\documentclass[11pt]{llncs}
\usepackage{a4wide}
\usepackage[latin1]{inputenc}
\usepackage{graphicx}
\usepackage{color}
\usepackage{amssymb}
\usepackage{setspace}

\usepackage[ruled]{algorithm2e}

\title{Post-Processing Hierarchical Community Structures:\\Quality Improvements and Multi-scale View}
\author{Pascal Pons and Matthieu Latapy}
\institute{LIP6 -- CNRS and Universit\'e Pierre et Marie Curie (UPMC -- Paris 6) \\ 104, avenue du Pr\'esident Kennedy, 75016 Paris, France \\ \email{Matthieu.Latapy@lip6.fr}}

\begin{document}
\maketitle
\begin{abstract}
Dense sub-graphs of sparse graphs (\emph{communities}), which appear in most real-world complex networks, play an important role in many contexts. Most existing community detection algorithms produce a hierarchical structure of communities and seek a partition into communities that optimizes a given quality function. We propose new methods to improve the results of any of these algorithms. First we show how to optimize a general class of additive quality functions (containing the modularity, the performance, and a new similarity-based quality function which we propose) over a larger set of partitions than the classical methods. Moreover, we define new multi-scale quality functions which make it possible to detect different scales at which meaningful community structures appear, while classical approaches find only one partition.
\end{abstract}


\section{Introduction}

Recent advances have emphasized the importance of \emph{complex networks} in many different domains such as sociology (acquaintance networks, collaboration networks), biology (metabolic networks, gene networks) or computer science (Internet topology, web graph, p2p networks, e-mail exchanges). We refer the reader to \cite{Brandes:Network_Analysis,Strogatz:2001,albert:2002,Newman:2003,Dorogovtsev:2003} for reviews from different perspectives and for an extensive bibliography.

%\subsection{Community detection}

 The analysis of these networks has brought out important and challenging graph algorithm problems. One of them is \emph{community detection}, used to uncover structure in large networks: the corresponding graphs are generally globally sparse but locally dense; there exist groups of vertices, called \emph{communities}, with many links between them but few links to other vertices. 
 Formally, we consider an undirected graph $G = (V, E)$ with $n = |V|$ vertices, $m = |E|$ edges. The aim of a community detection algorithm is to find a partition $\mathcal{P} = \{\mathcal{C}_1, \dots, \mathcal{C}_k\}$ of the vertices ($\mathcal{C}_i \cap \mathcal{C}_j = \emptyset$ for $i \neq j$ and $\cup_{i}\ \mathcal{C}_i = V$) that maximizes a given \emph{quality function} $Q(\mathcal{P})$ (see Section \ref{improving}).

Various approaches exist; they belong to a few main methodological categories which we succinctly overview here. First, the \emph{divisive} approach starts from the entire graph and successively splits it into more and more communities. Some algorithms achieve this by removing inter-communities edges (the communities are the remaining connected components) according to their betweenness \cite{Newman_Girvan:2004,Fortunato:2004} or their local clustering \cite{Radicchi_Filippo:2004}. Others use recursive bisection mechanisms based on minimum cuts \cite{hartuv:00_clustering} or spectral methods \cite{Kannan:2000}. Another family of approaches, the \emph{agglomerative} one, starts from $n$ single-vertex communities and merges them successively into larger and larger communities. Some algorithms use hierarchical clustering methods according to different similarity measurements based on spectral properties \cite{Donetti:2004} or random walks \cite{pons_latapy:2005,pons_latapy:2006,ZhouL04}. Other algorithms are based on greedy optimization of a quality function \cite{Blondel2008fast,Clauset_Newman:2004}. Finally, \emph{direct} approaches trying to perform global optimization of a quality function \cite{Duch_Arenas:2005,Guimera_Amaral:2005}, or iteratively modifying the weight of the edges to make clusters appear \cite{mcl}, have also been used.

Most community detection algorithms induce series of partitions $\mathcal{P}_0, \dots, \mathcal{P}_c$ corresponding to successive steps of the algorithm: $\mathcal{P}_{k+1} = \mathcal{P}_k\backslash\{\mathcal{C}_k\} \cup \{\mathcal{C}_1', \dots, \mathcal{C}_j'\}$ with $\mathcal{C}_k = \cup_{i=1}^j\mathcal{C}_i'$ and $\mathcal{P}_0 = V$. If one considers a divisive algorithm the partitions $\mathcal{P}_k$ are obtained in increasing order of the steps $k$, and in decreasing order if one considers an agglomerative algorithm. Classical community detection algorithms output the partition that maximizes a given quality function $Q$ among the $c$ partitions $\mathcal{P}_0, \dots, \mathcal{P}_c$.

One then defines the {\em dendrogram} associated to the running of the algorithm as the tree in which $\mathcal{C}_1', \dots, \mathcal{C}_j'$ are the children of $\mathcal{C}_k$, with the above notation, for all step $k$. We consider in this paper the situation where the dendrogram resulting from a running of a community detection algorithm on $G$ is given. There are at most $n$ steps as described above ($c < n$), which produce a set $S$ of $c+n$ subsets of $V$: $c$ subsets $\mathcal{C}_k$ corresponding to the steps of the algorithm plus $n$ single-vertex sets. Many possible partitions are induced by these subsets; we denote $\Pi$ the set of all these possible partitions: $\Pi = \{\mathcal{P}| \forall \mathcal{C} \in \mathcal{P}, \mathcal{C} \in S \textrm{ and } \cup_{\mathcal{C}\in \mathcal{P}} \mathcal{C} = V \textrm{ and } \forall \mathcal{C}_i \neq \mathcal{C}_j \in \mathcal{P}, \mathcal{C}_i \cap \mathcal{C}_j = \emptyset\}$. Intuitively these partitions are given by horizontal (but not necessarily straight) cuts of the associated dendrogram (Figure~\ref{figure:example}b). We also define in the same manner the sets $\Pi_{\mathcal{C}}$ of all possible partitions of a community $\mathcal{C}$ in $S$. The reader must keep in mind that, throughout this paper, we will never consider any partition (or sub-partition) containing a community that is not in $S$, the set of all communities induced by the given dendrogram.

\subsection*{Contribution}

We introduce in this paper new post-processing methods to improve the results of any algorithm that finds hierarchical community structures (encoded by the dendrogram). We address the two following limitations of previous contributions.
 
First, we note that considering all possible partitions in $\Pi$ (instead of only the $c+1$ partitions $\mathcal{P}_0, \dots, \mathcal{P}_c$) will necessarily produce better results than the classical method, and cannot be worse. The number of valid partitions being exponential in general, it is impossible to find efficiently the partition that maximizes an arbitrary quality function. However, we will show in Section~\ref{improving} that this is possible with some reasonable assumptions on the quality function $Q(\mathcal{P})$. These results are obtained for a general class of \emph{additive quality functions} that contains the \emph{modularity} \cite{Newman_Girvan:2004}, the \emph{performance} \cite{Brandes:Network_Analysis} and a new similarity-based quality function which we introduce in Section~\ref{improving}.

Second, we propose in Section~\ref{multi-scale} \emph{multi-scale quality functions} in order to detect community structures at different scales and to determine the most relevant scales at which the graph should be observed.

We will finally evaluate the benefits of these new approaches with some experiments (Section~\ref{experiments}).

\section{Improving the partition into communities} \label{improving}

In this section, we first introduce a general class of additive quality functions. We show that such functions can be efficiently optimized\footnote{Without loss of generality we can consider that the function must be maximized.} over all possible partitions $\mathcal{P} \in \Pi$ encoded in a dendrogram.

\begin{definition} \label{def:additivity} A quality function $Q$ is \textbf{additive} if there exists an elementary function $q$, such that for any partition $\mathcal{P}$: 
$$Q(\mathcal{P}) = \sum_{\mathcal{C} \in \mathcal{P}} q(\mathcal{C})$$ 
%Its restriction $Q^{\mathcal{C}}$ to any sub-partition $\mathcal{P}^{\mathcal{C}}$ of the vertices of a community $\mathcal{C}$ is:
%$$Q^{\mathcal{C}}(\mathcal{P}^{\mathcal{C}}) = \sum_{\bar{\mathcal{C}} \in \mathcal{P}^{\mathcal{C}}} q(\bar{\mathcal{C}})$$ 
\end{definition}

\noindent 
Let us first show that this definition is not too restrictive by considering three special cases of interest.

\paragraph{The modularity} introduced in \cite{Newman_Girvan:2004} has already been widely used \cite{Blondel2008fast,Clauset_Newman:2004,Donetti:2004,Duch_Arenas:2005,Fortunato:2004,Guimera_Amaral:2005,Newman_Girvan:2004}. It relies on the internal and total fractions of edges bond to a community $\mathcal{C}$, respectively $e(\mathcal{C}) = \displaystyle\sum_{i \in \mathcal{C}} \sum_{j \in \mathcal{C}} \frac{A_{ij}}{2m}$ and $a(\mathcal{C}) = \displaystyle\sum_{i \in \mathcal{C}} \sum_{j \in V} \frac{A_{ij}}{2m}$ ($A$ is the adjacency matrix and $m$ the number of edges).
$$
Q^M(\mathcal{P}) = \sum_{\mathcal{C} \in \mathcal{P}} e(\mathcal{C}) - a(\mathcal{C})^2
$$
This definition directly induces that the modularity is additive, using $q^M(\mathcal{C}) = e(\mathcal{C}) - a(\mathcal{C})^2$. We may also notice that it satisfies $-1 \leq Q^M(\mathcal{P}) \leq 1$, and that each evaluation of the function can be done in $\mathcal{O}(m)$.

\paragraph{The performance} \cite{Brandes:Network_Analysis} counts the number of correctly classified pairs of vertices (either two vertices belonging to the same community and connected by an edge, or two vertices belonging to different communities and not connected by an edge): 
$$
Q^P(\mathcal{P}) = \frac{|\{\{u,v\} \in E, \mathcal{C}(u) = \mathcal{C}(v)\}| + |\{\{u,v\} \notin E, \mathcal{C}(u) \neq \mathcal{C}(v)\}|}{\frac{1}{2}n(n-1)}
$$
where $\mathcal{C}(u)$ denotes the community containing vertex $u$ in the partition $\mathcal{P}$. The function $Q^P$ is the fraction of correctly identified pairs of links, and so $0 \leq Q^P(\mathcal{P}) \leq 1$. Its additivity is proved using $q^P(\mathcal{C}) = \frac{1}{n(n-1)}\sum_{u \in \mathcal{C}} |\{v \in \mathcal{C}, \{u,v\} \in E\}| + |\{v \notin \mathcal{C}, \{u,v\} \notin E\}|$. This quality function can be computed in $\mathcal{O}(n^2)$ and may be generalized to weighted graph as discussed in \cite{Brandes:Network_Analysis}.

\paragraph{A similarity based quality function.}
This approach supposes that we have a distance $d_{ij} \geq 0$ measuring the similarity between any pair of vertices $i$ and $j$ (the smaller $d_{ij}$ is, the more similar $i$ and $j$ are). We want to find homogeneous communities by minimizing their heterogeneity quantified by the mean square sum of the distances $\sigma(\mathcal{C}) = \frac{1}{|\mathcal{C}|}\sum_{i,j \in \mathcal{C}} d_{ij}^2$.

However, minimizing these quantities leads to the partition with $n$ single-vertex community. We will avoid this by minimizing at the same time the number $c(\mathcal{P})$ of communities in the partition. The maximal values of these quantities (namely $\sigma(\mathcal{C}) \leq \sigma_{max}$ obtained for $\mathcal{C} = V$, and $c(\mathcal{P}) \leq n$) are used in the following definition:
$$
Q^S(\mathcal{P}) =  -\frac{c(\mathcal{P})}{n} - \sum_{\mathcal{C} \in \mathcal{P}} \frac{\sigma(\mathcal{C})}{\sigma_{max}}
$$
This quality function satisfies $-2 \leq Q^S(\mathcal{P}) \leq 0$. We prove that it is additive using $q^S(\mathcal{C}) = -\frac{1}{n} - \frac{\sigma(\mathcal{C})}{\sigma_{max}}$. Each evaluation of $\sigma(\mathcal{C})$ requires $\mathcal{O}(|\mathcal{C}|^2)$ distance computations for an arbitrary distance. However if $d$ is an Euclidean distance then $\sigma(\mathcal{C}_i \cup \mathcal{C}_j)$ can be obtained from $\sigma(\mathcal{C}_i)$ and $\sigma(\mathcal{C}_j)$ with only one additional distance computation. Therefore all the $\sigma(\mathcal{C})$ can be obtained with $n$ distance computations in this case. Such a distance, based on random walks, was proposed in \cite{pons_latapy:2005,pons_latapy:2006} together with an agglomerative community detection algorithm which computes the values of $\sigma(\mathcal{C})$. Thus this quality function can be used within the framework presented here at no additional cost.

\bigskip

The examples above show that the class of additive quality functions is quite general, and that many previously used quality functions actually fit in this class. We will now show that it is possible to maximize any additive quality function over the set of partitions $\Pi$ with a simple recursive approach.

\begin{lemma} \label{lemma1} Given an additive quality function $Q$ and a dendrogram in which the set $\mathcal{C}$ has children $\mathcal{C}_1, ..., \mathcal{C}_k$, the partition $\mathcal{P}_{max} \in \Pi_{\mathcal{C}}$ that maximizes $Q$ is either $\{\mathcal{C}\}$ or $\mathcal{P}_1 \cup ... \cup \mathcal{P}_k$ where $\mathcal{P}_i \in \Pi_{\mathcal{C}_i}$ maximizes $Q$ over $\Pi_{\mathcal{C}_i}$ and $Q(\mathcal{P}_{max}) = \displaystyle\max_{\mathcal{P} \in \Pi_{\mathcal{C}}} Q(\mathcal{P}) = \max(q(\mathcal{C}), \sum_i Q(\mathcal{P}_i))$.
\end{lemma}
\begin{proof}
Suppose that the partition $\mathcal{P}_{max} \in \Pi_{\mathcal{C}}$ maximizing $Q$ is not $\{\mathcal{C}\}$. Then it induces a sub-partition $\mathcal{P}_i \in \Pi_{\mathcal{C}_i}$ in each of its children $\mathcal{C}_i$ such that $\mathcal{P}_{max} = \cup_i \mathcal{P}_i$. Now suppose there exists $\mathcal{P}_i' \in \Pi_{\mathcal{C}_i}$ such that $Q(\mathcal{P}_i') > Q(\mathcal{P}_i)$. Then the sub-partition $\mathcal{P}_{max}' = \mathcal{P}_1 \cup ...\cup \mathcal{P}_i' \cup \dots \cup \mathcal{P}_k$ will satisfy, thanks to additivity, $Q(\mathcal{P}_{max}') > Q(\mathcal{P}_{max})$, which is impossible.
\qed\end{proof}

\begin{theorem} Given an additive quality function $Q$ and a dendrogram, it is possible to find the partition $\mathcal{P} \in \Pi$ that maximizes $Q$ with $\mathcal{O}(n)$ evaluations of function $q$. This is achieved by function \verb#FindBestPartition#.
\end{theorem}
\begin{proof}
Lemma~\ref{lemma1} guarantees that the recursive function FindBestPartition finds the partition maximizing $Q$ over $\Pi$ when called on the largest set of vertices $V$. The function is called only once on each node of the dendrogram, thus the total number of calls (and thus the total number of evaluations of the elementary quality function $q$) is $|S| \leq 2n$.
\qed\end{proof}

%\vspace{-8mm}
\begin{function}
\SetKwFunction{FindBestPartition}{FindBestPartition}
\caption{FindBestPartition($\mathcal{C}$)}

\ForEach{child $\mathcal{C}_i$ of $\mathcal{C}$}
	{($Q_i, \mathcal{P}_i$) $\leftarrow$ \FindBestPartition{$\mathcal{C}_i$}}
\uIf{$\mathcal{C}$ has no child or $q(\mathcal{C}) > \sum_i Q_i$}
	{\Return{$q(\mathcal{C}),\{\mathcal{C}\}$}}
\uElse{\Return{$\sum_i Q_i,\cup_i \mathcal{P}_i$}}
\end{function}
%\vspace{-7mm}

Let us note moreover that some quality functions allow optimizations concerning the computation of the $q(\mathcal{C})$: for example it is possible to compute efficiently $q(\mathcal{C}_i \cup \mathcal{C}_j)$ from the values of $q(\mathcal{C}_i)$ and $q(\mathcal{C}_j)$ for the modularity \cite{Clauset_Newman:2004} and for the random walk quality function \cite{pons_latapy:2005,pons_latapy:2006}.

\begin{figure}[!h]
\begin{center}
\includegraphics[width=\textwidth]{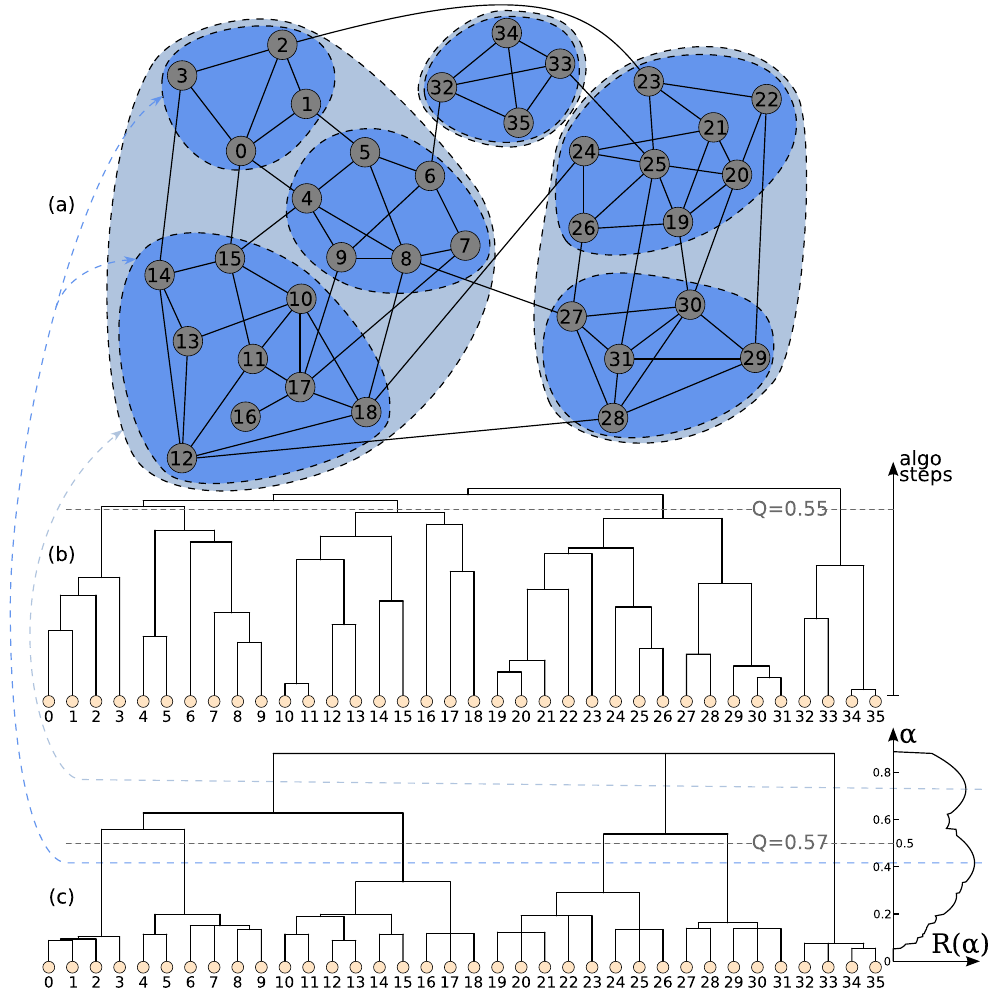}
\caption{{\bf(a)} Example graph with a multi-scale community structure. {\bf(b)} Hierarchical community structure (dendrogram) found by the Walktrap algorithm \cite{pons_latapy:2005,pons_latapy:2006}: the heights of the nodes represent the steps of the algorithm. The classical approach only considers partitions given by straight horizontal cuts on this dendrogram: here a partition into 5 communities maximizes the modularity $Q^M = 0.55$. {\bf(c)} Reordered dendrogram according the multi-scale quality function $Q_{\alpha}^M$. Horizontal cuts show the best partition $\mathcal{P}_{\alpha}$ for any scale factor $\alpha$. The maximal modularity $Q^M = 0.57$ (obtained for $\alpha = \frac{1}{2}$) improves the classical approach by finding a better partition in the dendrogram. In addition, the relevance function $R(\alpha)$ indicates two meaningful scale factors ($\alpha = 0.42$ and $\alpha = 0.73$) corresponding to a partition into 6 communities and a partition into 3 communities (outlined in dark blue and light blue respectively). Notice moreover that these partitions are obtained for wide ranges of values of $\alpha$, which may be seen as an indication of the fact that they are very relevant.}
%\vspace{0.5cm}
\label{figure:example}
\end{center}
\end{figure}

\clearpage

\section{Multi-scale community structure detection} \label{multi-scale}
Even if most community detection algorithms find hierarchical community structures, they generally ouptput only one partition (like in Section~\ref{improving}). However, communities often appear at different scales in complex networks. To overcome this limitation, we will propose here multi-scale quality functions which work at different scales. We will then propose a method to determine the most relevant scales, highlighting meaningful communities.

\subsection{Multi-scale quality functions} \label{multi-scale_function}

We will consider in this section a scale factor $0 \leq \alpha \leq 1$ going from microscopic to macroscopic scales: $\alpha = 0$ corresponds to smallest communities with only one vertex and $\alpha = 1$ corresponds to the largest community containing all the vertices. We will define multi-scale quality functions $Q_{\alpha}$ and the partitions $\mathcal{P}_{\alpha}$ maximizing them should be consistent with the scale factor, which is captured by the following definition.

\begin{definition} \label{def:multi-scale}
Consider a family of quality functions $(Q_{\alpha})_{0 \leq \alpha \leq 1}$, and denote by $\mathcal{P}_{\alpha}$ the partition in $\Pi$ maximizing $Q_{\alpha}$. Then $(Q_{\alpha})_{0 \leq \alpha \leq 1}$ are \textbf{multi-scale quality functions} if 
$$
\alpha_1 \leq \alpha_2 \Rightarrow \mathcal{P}_{\alpha_1} \preceq \mathcal{P}_{\alpha_2} \textrm{ with } \mathcal{P}_{\alpha = 0} = \{\{v\}|v \in V\} \textrm{ and } \mathcal{P}_{\alpha = 1} = \{V\}
$$
where $\mathcal{P}_{\alpha_1} \preceq \mathcal{P}_{\alpha_2}$ iff $\mathcal{P}_{\alpha_1}$ is a refinement of $\mathcal{P}_{\alpha_2}$, {\em i.e.} the sets of $\mathcal{P}_{\alpha_1}$ are included in those of $\mathcal{P}_{\alpha_2}$: for all $\mathcal{C}_1 \in \mathcal{P}_{\alpha_1}$, there exists $\mathcal{C}_2 \in \mathcal{P}_{\alpha_2}$ such that $\mathcal{C}_1 \subseteq \mathcal{C}_2$
\end{definition}
Note that for any $\alpha$, $Q_{\alpha}$ is a quality function, and so the notion of additivity (Definition~\ref{def:additivity}) applies. We propose now a general class of additive multi-scale quality functions.
\begin{theorem} \label{th_general_multiscale}

Let us consider a function $h$ over the parts of $V$ defined by a given dendrogram, such that $h(\mathcal{C}_i \cup \mathcal{C}_j) \geq h(\mathcal{C}_i) + h(\mathcal{C}_j)$: $h$ is larger in macroscopic scales. Likewise, let us consider $l$ such that $l(\mathcal{C}_i \cup \mathcal{C}_j) \leq l(\mathcal{C}_i) + l(\mathcal{C}_j)$: $l$ is larger in microscopic scales. Then functions $Q_{\alpha}$ defined by:
$$
Q_{\alpha}(\mathcal{P}) = \sum_{\mathcal{C} \in \mathcal{P}} q_{\alpha}(\mathcal{C}) \textrm{ with } q_{\alpha}(\mathcal{C}) = \alpha h(\mathcal{C}) + (1-\alpha)l(\mathcal{C})
$$
are additive multi-scale quality functions.
\end{theorem}

\begin{proof} Suppose that $\alpha_1 < \alpha_2$ but $\mathcal{P}_{\alpha_1} \npreceq \mathcal{P}_{\alpha_2}$. Then there exist $\mathcal{C} \in \mathcal{P}_{\alpha_1}$ and $\mathcal{C}_1, \dots, \mathcal{C}_k  \in \mathcal{P}_{\alpha_2}$ such that $\mathcal{C} = \mathcal{C}_1 \cup \dots \cup \mathcal{C}_k$. We have: $q_{\alpha_1}(\mathcal{C}) = \alpha_1 h(\mathcal{C}) + (1 - \alpha_1)l(\mathcal{C}) = q_{\alpha_2}(\mathcal{C}) + (\alpha_1 - \alpha_2)h(\mathcal{C}) + (\alpha_2 - \alpha_1)l(\mathcal{C})$. But $\mathcal{P}_{\alpha_2}$ (containing $\mathcal{C}_1, \dots, \mathcal{C}_k$) maximizes $Q_{\alpha_2}$, therefore: $q_{\alpha_2}(\mathcal{C}) \leq q_{\alpha_2}(\mathcal{C}_1) + \dots + q_{\alpha_2}(\mathcal{C}_k)$. Moreover $h$ and $l$ satisfy: $h(\mathcal{C}) \geq h(\mathcal{C}_1) + \dots + h(\mathcal{C}_k)$ and $l(\mathcal{C}) \leq l(\mathcal{C}_1) + \dots + l(\mathcal{C}_k)$. Finally with the fact that $\alpha_1 < \alpha_2$ we obtain: $q_{\alpha_1}(\mathcal{C}) \leq q_{\alpha_2}(\mathcal{C}_1) + \dots + q_{\alpha_2}(\mathcal{C}_k) + (\alpha_1 - \alpha_2)(h(\mathcal{C}_1) + \dots + h(\mathcal{C}_k)) + (\alpha_2 - \alpha_1)(l(\mathcal{C}_1) + \dots + l(\mathcal{C}_k))$. We recognize the inequality $q_{\alpha_1}(\mathcal{C}) \leq q_{\alpha_1}(\mathcal{C}_1) + \dots +q_{\alpha_1}(\mathcal{C}_k)$ which is in contradiction with the fact that $\mathcal{P}_{\alpha_1}$ maximizes $Q_{\alpha_1}$. This proves the main property of Definition~\ref{def:multi-scale}.\\
Then the additivity is immediate and it is simple to check that $\mathcal{P}_{\alpha = 0} = \{\{v\}|v \in V\}$ and $\mathcal{P}_{\alpha = 1} = \{V\}$ thanks to the inequalities satisfied by $h$ and $l$.
\qed\end{proof}

This theorem makes it possible to create an additive multi-scale quality function from two elementary functions. These two functions must have opposite growing behavior with community sizes and they also have to capture expected properties of communities. We now propose suitable multi-scale quality functions which generalize those of Section~\ref{improving} (the original quality functions are obtained back as a particular case for $\alpha = \frac{1}{2}$).

\paragraph{The multi-scale modularity.}
We generalize the modularity by introducing the scale factor $\alpha$ in its definition:
$$
Q_{\alpha}^M(\mathcal{P}) = \sum_{\mathcal{C} \in \mathcal{P}} \alpha e(\mathcal{C}) - (1-\alpha)a(\mathcal{C})^2
$$
We check that the properties of Theorem~\ref{th_general_multiscale} are satisfied to ensure that $Q_{\alpha}^M$ is an additive multi-scale quality function. We consider $h^M(\mathcal{C}) = e(\mathcal{C})$ the fraction of internal edges of community $\mathcal{C}$ and $l^M(\mathcal{C}) = -a(\mathcal{C})^2$ using the fraction of edges bound to community $\mathcal{C}$. We have $h^M(\mathcal{C}_i \cup \mathcal{C}_j) \geq h^M(\mathcal{C}_i) + h^M(\mathcal{C}_j)$ because $e(\mathcal{C}_i \cup \mathcal{C}_j) = e(\mathcal{C}_i) + e(\mathcal{C}_j) + (\mbox{fraction}\ \mbox{of}\ \mbox{edges}\ \mbox{between}\ \mathcal{C}_i \textrm{ and } \mathcal{C}_j)$. And $l^M(\mathcal{C}_i \cup \mathcal{C}_j) \leq l^M(\mathcal{C}_i) + l^M(\mathcal{C}_j)$ because $a(\mathcal{C}_i \cup \mathcal{C}_j) = a(\mathcal{C}_i) + a(\mathcal{C}_j)$ and thus $l^M(\mathcal{C}_i) + l^M(\mathcal{C}_j) - l^M(\mathcal{C}_i \cup \mathcal{C}_j) = 2a(\mathcal{C}_i)a(\mathcal{C}_j)$.

\paragraph{The multi-scale performance.} It is defined in the same manner by:
$$
Q^P_{\alpha}(\mathcal{P}) = \frac{\alpha |\{\{u,v\} \in E, \mathcal{C}(u) = \mathcal{C}(v)\}| + (1-\alpha) |\{\{u,v\} \notin E, \mathcal{C}(u) \neq \mathcal{C}(v)\}|}{\frac{1}{2}n(n-1)}
$$
We use $h^P(\mathcal{C}) = \frac{1}{n(n-1)}\sum_{u \in \mathcal{C}} |\{v \in \mathcal{C}, \{u,v\} \in E\}|$ and $l^P(\mathcal{C}) = \frac{1}{n(n-1)}\sum_{u \in \mathcal{C}} |\{v \notin \mathcal{C}, \{u,v\} \notin E\}|$. The two inequalities required by Theorem~\ref{th_general_multiscale} are easily verified if we remark that $h^P(\mathcal{C})$ counts the number of edges inside $\mathcal{C}$ and $l^P(\mathcal{C})$ counts the number non existing edges between vertices of $\mathcal{C}$ and other vertices.

\paragraph{A multi-scale similarity based quality function.} Using the same idea we can generalize the third quality function based on similarity measurement $d_{ij}$ between vertices. However, the quantity $\sigma(\mathcal{C})$ measuring community homogeneity must also satisfy $\sigma(\mathcal{C}_i \cup \mathcal{C}_j) \geq \sigma(\mathcal{C}_i) + \sigma(\mathcal{C}_j)$, which is the case for Euclidean distances.
$$
Q^S_{\alpha}(\mathcal{P}) = -\alpha c(\mathcal{P}) - (1-\alpha) \sum_{\mathcal{C} \in \mathcal{P}} \frac{\sigma(\mathcal{C})}{\sigma_{max}}
$$
$h^S(\mathcal{C}) = -\frac{1}{n}$ trivially satisfies the inequality of Theorem~\ref{th_general_multiscale}. The other inequality satisfied by $l^S(\mathcal{C}) = -\frac{\sigma(\mathcal{C})}{\sigma_{max}}$ comes from the restriction on $d$ pointed out above.

\subsection{Finding the best partition for every scale} \label{find_all_best_partitions}

A multi-scale quality function $Q_{\alpha}$ allows us to find a partition $\mathcal{P}_{\alpha}$ for any scale factor $0 \leq \alpha \leq 1$. We will show in this section how to compute efficiently all these partitions for the general class of multi-scale quality functions defined in Theorem~\ref{th_general_multiscale}. The order  between the $\mathcal{P}_{\alpha}$ (Definition~\ref{def:multi-scale} indicates that $\alpha_1 \leq \alpha_2 \Rightarrow \mathcal{P}_{\alpha_1} \preceq \mathcal{P}_{\alpha_2}$) implies that the total number of different partitions $\mathcal{P}_{\alpha}$ is at most $n$. Indeed, each partition is obtained from the previous one by splitting at least one community. Therefore the number of communities of the $k^{\textrm{\small{th}}}$ partition is at least $k$. The number of communities of each partition being less than $n$ (the number of vertices) we cannot have more than $n$ different partitions $\mathcal{P}_{\alpha}$.

To determine all the partitions $\mathcal{P}_{\alpha}$, we only need to determine the list of the particular scale factors $\alpha_i$ at which $\mathcal{P}_{\alpha}$ changes (split of a community into sub-communities). The corresponding modifications induce a new hierarchy into the community structure: the community splits can be ordered by scale factors $\alpha_i$ at which they occur. The dendrogram can be reordered with this new hierarchy as illustrated in Figure~\ref{figure:example}c. This provides more accurate information on community scales and improves comparison between them.

For each partition $\mathcal{P}$, the function $Q_{\alpha}(\mathcal{P}) = l(\mathcal{P}) + (h(\mathcal{P}) - l(\mathcal{P}))\alpha$ can be seen as an affine function of the parameter $\alpha$. Therefore, finding all the best partitions $\mathcal{P}_{\alpha}$ is equivalent to finding the function $Q^{\Pi}_{max}(\alpha) = Q_{\alpha}(\mathcal{P}_{\alpha})$ defined as follows.

\begin{definition} The piecewise affine function $Q^{\Pi_{\mathcal{C}}}_{max}(\alpha)$ maximizes $Q_{\alpha}(\mathcal{P})$ over all possible partitions $\mathcal{P} \in \Pi_{\mathcal{C}}$:
$$
Q^{\Pi_{\mathcal{C}}}_{max}(\alpha) = \max_{\mathcal{P} \in \Pi_{\mathcal{C}}} Q_{\alpha}(\mathcal{P})
$$ 
\end{definition}

\begin{theorem} Given additive multi-scale quality functions $Q_{\alpha}$ satisfying Theorem~\ref{th_general_multiscale} and a dendrogram, it is possible to compute $Q^{\Pi}_{max}(\alpha)$ by making at most $\mathcal{O}(n)$ evaluations of the elementary quality function $q_{\alpha}$. The additional average complexity is $\mathcal{O}(n\sqrt{n})$ for an arbitrary dendrogram, it is $\mathcal{O}(n\log(n))$ for balanced ones and the worst case is $\mathcal{O}(n^2)$. This is achieved by the function \verb#FindMultiscalePartitions#.
\end{theorem}

\begin{proof}
For a given $\alpha$, and for a community $\mathcal{C}$ having children $\mathcal{C}_1, \dots, \mathcal{C}_k$ in the dendrogram, Lemma~\ref{lemma1} indicates that ${\displaystyle \max_{\mathcal{P} \in \Pi_{\mathcal{C}}}} Q_{\alpha}(\mathcal{P}) = \max(q_{\alpha}(\mathcal{C}), \sum_i {\displaystyle \max_{\mathcal{P} \in \Pi_{\mathcal{C}_i}}} Q_{\alpha}(\mathcal{P}))$. This equality holds for any $\alpha$ and thus we deduce $Q^{\Pi_{\mathcal{C}}}_{max}(\alpha) = \max(q_{\alpha}(\mathcal{C}), \sum_i Q^{\Pi_{\mathcal{C}_i}}_{max}(\alpha))$. This proves the corectness of the recursive function FindMultiscalePartitions that computes $Q_{max}^{\Pi_{\mathcal{C}}}$ by manipulating piecewise affine functions. $Q^{\Pi}_{max}(\alpha)$ is obtained for parameter $\mathcal{C} = V$.

The function is recursively called exactly once on each node of the dendrogram, leading to $\mathcal{O}(n)$ evaluations of the elementary quality function $q_{\alpha}$. Each call also evaluates a sum and a maximum operation on piecewise affine functions encoded by the list of their particular points $(\alpha_i, Q^{\Pi_{\mathcal{C}_i}}_{max}(\alpha_i))$. These operations are done in time linear in the size of input piecewise functions, and the sum of their sizes is at most $|\mathcal{C}|$. Therefore this additional complexity is represented by the sum over all the nodes of the dendrogram of operations in $\mathcal{O}(|\mathcal{C}|)$. We can notice that this sum is nothing else than the path length of the hierarchical tree structure of community. Classical analysis shows that the path length is between $n\log(n)$ and $n^2$ with an average value (over all trees of size $n$) in $\mathcal{O}(n\sqrt{n})$ \cite{SeFl96a}.
\qed\end{proof}

%\vspace{-8mm}
\begin{function}
\SetKwFunction{FindMultiscalePartitions}{FindMultiscalePartitions}
\caption{FindMultiscalePartitions($\mathcal{C}$)}

\ForEach{child $\mathcal{C}_i$ of $\mathcal{C}$}
	{$Q^{\Pi_{\mathcal{C}_i}}_{max} \leftarrow$ \FindMultiscalePartitions{$\mathcal{C}_i$}\;}
\uIf{$\mathcal{C}$ has no child}
	{\Return{$\alpha \mapsto q_{\alpha}(\mathcal{C})$}}
\uElse{
	\Return{$\alpha \mapsto \max(q_{\alpha}(\mathcal{C}), \sum_i Q^{\Pi_{\mathcal{C}_i}}_{max})$}
}
\end{function}
%\vspace{-7mm}

During the computation, we can keep in memory the communities $\mathcal{C}_i$ that are split at each scale factor $\alpha_i$. This provides all necessary information to know at which scale factor $\alpha$ each community appears and disappears from the partitions $\mathcal{P}_{\alpha}$. This also makes it possible to build the reorganized dendrogram and all partitions $\mathcal{P}_{\alpha}$ (see Figure~\ref{figure:example}c). 

If we compare the complexity of this post-processing algorithm to those of the known community detection algorithms, we can deduce that it may be integrated after almost any of them without changing their overall complexity. Moreover, hierarchical structures obtained from real cases tend to be balanced \cite{Clauset_Newman:2004}, which is the most favorable case for our complexity.

\subsection{A notion of scale relevance}

We showed that one can obtain all best partitions $\mathcal{P}_{\alpha}$ for any scale factor $\alpha$. However all these partitions may not have the same relevance in term of community structure. We will provide in this section a method to estimate the relevance of these partitions and to retrieve the most meaningful scale factors at which clear community structures appear.

The algorithm of Section~\ref{find_all_best_partitions} allows us to know when each community $\mathcal{C}$ appears and disappears from the partitions $\mathcal{P}_{\alpha}$. Let $\alpha_{min}(\mathcal{C})$ and $\alpha_{max}(\mathcal{C})$ be these two scale factors: $\mathcal{C} \in \mathcal{P}_{\alpha}$ for $\alpha_{min}(\mathcal{C}) < \alpha < \alpha_{max}(\mathcal{C})$. One may consider that the most relevant communities will be present for wide ranges of scale factors. We use this to measure the relevance of a community $\mathcal{C}$ by $\alpha_{max}(\mathcal{C}) - \alpha_{min}(\mathcal{C})$ and the best scale representing $\mathcal{C}$ as $\alpha = \frac{\alpha_{max}(\mathcal{C}) - \alpha_{min}(\mathcal{C})}{2}$. These two notions are captured by the following definition.
\begin{definition} We define the relevance function $R_{\alpha}(\mathcal{C})$ of a community $\mathcal{C}$ at scale $\alpha$ by:
$$
R_{\alpha}(\mathcal{C}) = \frac{\alpha_{max}(\mathcal{C}) - \alpha_{min}(\mathcal{C})}{2} + \frac{2(\alpha_{max}(\mathcal{C})-\alpha)(\alpha-\alpha_{min}(\mathcal{C}))}{\alpha_{max}(\mathcal{C}) - \alpha_{min}(\mathcal{C})}
$$
This leads to the global relevance function $\displaystyle R(\alpha) = \frac{1}{n}\sum_{\mathcal{C} \in \mathcal{P}_{\alpha}} |\mathcal{C}|R_{\alpha}(\mathcal{C})$.
%$$
%R(\alpha) = \sum_{\mathcal{C} \in \mathcal{P}_{\alpha}} \frac{|\mathcal{C}|R_{\alpha}(\mathcal{C})}{n} = \sum_{\mathcal{C} \in \mathcal{P}_{\alpha}} \frac{|\mathcal{C}|}{n} \Big(\frac{\alpha_{max}(\mathcal{C}) - \alpha_{min}(\mathcal{C})}{2} + \frac{2(\alpha_{max}(\mathcal{C})-\alpha)(\alpha-\alpha_{min}(\mathcal{C}))}{\alpha_{max}(\mathcal{C}) - \alpha_{min}(\mathcal{C})}\Big)
%$$

\end{definition}

$R_{\alpha}(\mathcal{C})$ is a quadratic function of $\alpha$. Its maximum is $R(\frac{\alpha_{max}(\mathcal{C}) - \alpha_{min}(\mathcal{C})}{2}) = \alpha_{max}(\mathcal{C}) - \alpha_{min}(\mathcal{C})$ and $R(\alpha_{min}(\mathcal{C})) = R(\alpha_{max}(\mathcal{C})) = \frac{\alpha_{max}(\mathcal{C}) - \alpha_{min}(\mathcal{C})}{2}$. It may be used for determining the scale factors corresponding to relevant community structures. We can use it to find the best scale $\alpha$ which maximizes $R(\alpha)$, but we can also focus on other local maxima of $R(\alpha)$ corresponding to other interesting scales. This method allows us to determine several relevant scales and thus several relevant partitions (see Figure~\ref{figure:example}c for an example). 

The computation of $R(\alpha)$ and its maxima can be done in $\mathcal{O}(n)$. $R(\alpha)$ is a quadratic function that can be written as $R(\alpha) = A \alpha^2 + B\alpha + C$ between each specific $\alpha_{i}$ (the $\alpha_{i}$ correspond to splits of communities in the hierarchy given by partitions $\mathcal{P}_{\alpha}$). At each split, the coefficients $A, B$ and $C$ are modified according to the coefficients of $R_{\alpha}(\mathcal{C})$ of the corresponding communities. The previous algorithm gives the list of these splits, which allows to compute coefficients $A, B$ and $C$ by updating them at each $\alpha_i$. Each community leads to two updates (in constant time) of the coefficients (one when it appears at $\alpha_{max}(\mathcal{C})$ and one when it disappears at $\alpha_{min}(\mathcal{C})$), thus the overall complexity is $\mathcal{O}(n)$.

\section{Experimental evaluation} \label{experiments}

\begin{figure}[h!]
%\vspace{-10mm}
\begin{center}
\includegraphics{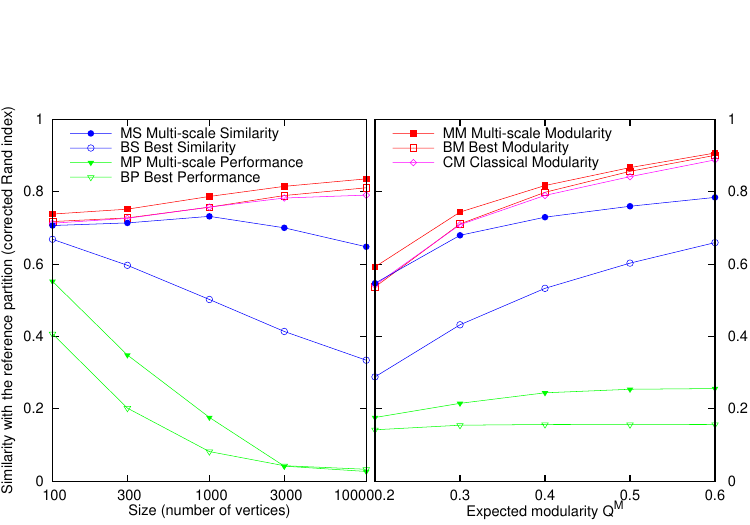}
\caption{Performance of the different methods measured by the similarity between the partition found and the actual generated partition. Left: influence of the size of the graph. Right: influence of the modularity of the reference partition.}
\label{figure:size}
\end{center}
%\vspace{-8mm}
\end{figure}

In this section we evaluate and compare the performances of the different methods and quality functions presented in this paper. Comparing community detection results is a difficult task because one needs some test graphs whose community structure is already known. A classical approach is to use randomly generated graphs with communities. We will compare the partitions obtained by post-processing the results of the same agglomerative algorithm \cite{pons_latapy:2005,pons_latapy:2006} on a large set of such graphs.

We generate test graphs according to the following parameters: number of vertices $n$, number of communities $c$, average internal and external\,\footnote{In this paper, for a given graph divided into communities, {\em internal} edges are the ones linking two vertices in a same community; {\em external} edges are the ones linking vertices in two different communities.} degrees $d_{in}$ and $d_{out}$. We divide the $n$ vertices into $c$ equal-sized sets then we draw each possible edge with probabilities $p_{in}$ or $p_{out}$ chosen according to $d_{in}$ and $d_{out}$. We evaluate found partitions by comparing them to the original generated partition. To achieve this, we use the Rand index corrected by Hubert and Arabie \cite{Rand,Hubert_Arabie} which evaluates the similarities between two partitions. The Rand index $I(\mathcal{P}_i,\mathcal{P}_j)$ is the ratio of pairs of vertices correlated by the partitions $\mathcal{P}_i$ and $\mathcal{P}_j$ (two vertices are correlated by the partitions $\mathcal{P}_i$ and $\mathcal{P}_j$ if they are classified in the same community or in different communities in the two partitions). The expected value of $I$ for a random partition is not zero. To avoid this, Hubert and Arabie proposed a corrected index that is also more sensitive: $I' = \frac{I - I_{exp}}{I_{max} - I_{exp}}$ where $I_{exp}$ is the expected value of $I$ for two random partitions with the same community size as $\mathcal{P}_i$ and $\mathcal{P}_j$.

\begin{figure}[h!]
%\vspace{-10mm}
\begin{center}
\includegraphics{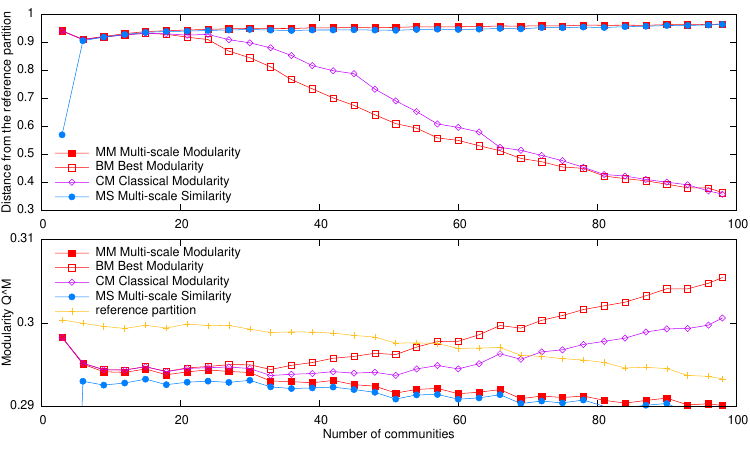}
\caption{Influence of the number of communities on generated graphs with $n = 1000$ vertices. Top: similarity between the partition found and the actual generated partition. Bottom: modularity $Q^M$ of the partition found and of the reference partition.}
\label{figure:nb_com}
\end{center}
%\vspace{-8mm}
\end{figure}

We will compare the following approaches: \emph{Classical Modularity} (CM) maximizes $Q^M$ over $\mathcal{P}_0, \dots, \mathcal{P}_{c}$; \emph{Best Modularity} (BM) maximizes $Q^M$ over $\Pi$; and \emph{Multi-scale Modularity} (MM) maximizes $Q_{\alpha}^M$ over $\Pi$ for the most relevant scale factor $\alpha$ given by $R(\alpha)$. Similarly, we define \emph{Best Performance} (BP) and \emph{Multi-scale Performance} (MP) using $Q^P$ and \emph{Best Similarity} (BS) and \emph{Multi-scale Similarity} (MS) using $Q^S$.

The first test considers a set of $25\,000$ graphs with different sizes ($100 \leq n \leq 10000$), different numbers of communities, different internal degrees $4 \leq d_{in} \leq 10$ and external degrees such that the expected modularity of the reference partition satisfies $0.2 \leq Q^M(\mathcal{P}_{ref}) \leq 0.6$. The results (Figure~\ref{figure:size}) show that the performance $Q^P$ is not very well suited for community detection in sparse networks because it gives too much importance to non-existing edges. We may also notice that the similarity based quality function $Q^S$ does not produce satisfying results without considering its multi-scale version $Q^S_{\alpha}$. And finally we see that the classical and the best modularity methods produce good results that are improved by the multi-scale approach.

The two next experiments show advantages of the multi-scale quality functions. First, we will test their ability to find communities at any scale by considering different sizes of communities. We generated a set of graphs with $n=1000$ vertices, the same internal degrees $d_{in} = 3$ and the same expected modularity $Q^M_{exp} = 0.3$, but they differ in their number of communities $2 \leq c \leq 100$. The results (Figure~\ref{figure:nb_com}) show that multi-scale approaches (MM and MS) find the good partition for any number and size of communities while CM and BM approaches have difficulties in finding small communities.

It is interesting to compare the value of modularity found by the different approaches. Of course the BM method obtains the largest value, but all other approaches find partitions that are more similar to the reference partition. Moreover, it shows that it is possible to find a bad partition (that does not represent the correct scale) with a larger modularity than the reference partition. This disadvantage of the modularity is addressed by the multi-scale modularity proposed in this paper.

\begin{figure}[h!]
\begin{center}
\includegraphics{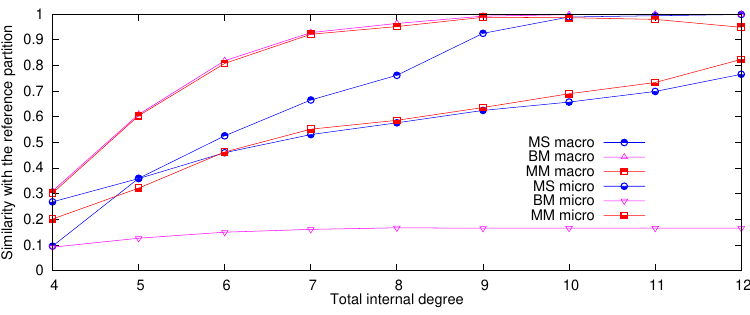}
\caption{Detection of communities at two different scales: distance from the macroscopic and the microscopic partitions in function of the total internal degree $d_{in} = d_{in}^{micro} + d_{in}^{macro}$.}
\label{figure:hierarchical}
\end{center}
%\vspace{-8mm}
\end{figure}

Finally we generated graphs with 1000 vertices and two community scales: vertices are divided into 10 communities that are themselves divided into 10 communities. This defines a macroscopic and a microscopic partition. Edges are randomly drawn in order to obtain three fixed average degrees $d_{in}^{micro}$, $d_{in}^{macro}$ and $d_{out}$ chosen between 2 and 6. We considered the two best scale factors indicated by the relevance function $R(\alpha)$ and compared the associated partitions to the two generated partitions. The results (Figure~\ref{figure:hierarchical}) show that the multi-scale quality functions make it possible to find distinct partitions corresponding to different scales. In comparison the BM method, that only find one partition, only detects the macroscopic partition.

\section{Conclusion}

We proposed in this paper methods improving the results of any community detection algorithm finding a hierarchical structure of communities. First, we showed how to optimize additive quality functions over a larger set of partitions than classical approaches. Moreover, we proposed multi-scale quality functions that work at different scales and make it possible to find more than only one relevant partition. Experiments have shown that these methods provide a significant improvement over classical approaches, especially in detecting small communities or communities that appear at different scales.

Moreover, scale factors associated with each community enable to reorder the dendrogram (Figure~\ref{figure:example}c), and we are convinced that they could also be integrated in a multi-scale visualization tool of complex networks based on community decomposition.

\section*{Acknowledgments}

We thank Cl\'emence Magnien for useful advice and helpful comments on preliminary versions. This work has been supported in part by the French national projects PERSI (Programme d'\'Etude des R\'eseaux Sociaux de l'Internet) and AGRI (Analyse des Grands R\'eseaux d'Interactions).

\bibliography{biblio}
\bibliographystyle{plain}

\end{document}